\documentclass[aps,pra,twocolumn,amsfonts,amssymb,amsmath,showpacs,showkeys,
floatfix,nofootinbib,groupedaddress,superscriptaddress,citesort]{revtex4}%
\usepackage{mathrsfs}
\usepackage{amsfonts}
\usepackage{amstext}
\usepackage{amsmath}
\usepackage{amssymb}
\usepackage{bm}
\usepackage{bbm}
\usepackage[dvips]{graphicx}
\def\qed{\leavevmode\unskip\penalty9999 \hbox{}\nobreak\hfill
     \quad\hbox{\leavevmode  \hbox to.77778em{%
              \hfil\vrule   \vbox to.675em%
               {\hrule width.6em\vfil\hrule}\vrule\hfil}}
     \par\vskip3pt}

\begin{document}

\preprint{APS/123-QED}
\title{Coherence measures and optimal conversion for coherent states}

\author{Shuanping Du}
\email{shuanpingdu@yahoo.com}\affiliation{School of Mathematical
Sciences, Xiamen University, Xiamen 361000,  China}

\author{Zhaofang Bai}\thanks{Corresponding author}
\email{baizhaofang@xmu.edu.cn}\affiliation{School of Mathematical
Sciences, Xiamen University, Xiamen 361000, China}

\author{Xiaofei Qi} \email{xiaofeiqisxu@aliyun.com}\affiliation{Department of Mathematics, Shanxi
University, Taiyuan 030006, China}


\begin{abstract}

We discuss a general strategy to construct coherence measures. One
can build an important class of coherence measures which cover the
relative entropy measure for pure states, the $l_1$-norm measure
for pure states and the $\alpha$-entropy measure.
 The optimal conversion of coherent
states under incoherent operations is presented which sheds some
light on the coherence of a single copy of a pure state.

\end{abstract}
\maketitle

\section{Introduction}
Superposition is a critical property of quantum system resulting
in quantum coherence and quantum entanglement. Quantum coherence
and also entanglement provide the important resource for quantum
information processing, for example, Deutsch¡¯s algorithm, Shor¡¯s
algorithm, teleportation, superdense coding and quantum
cryptography \cite{Nielsen}.

As with any such resource, there arises naturally the question of
how it can be quantified and manipulated. Attempts have been made
to find meaningful measures of entanglement
\cite{Ben1,Ben2,Ben3,Ved1,Ved2,vid2}, and also to uncover the
fundamental laws of its behavior under local quantum operations
and classical communication (LOCC)
\cite{Ben1,Ben2,Ben3,Ved1,Ved2,vid2,Nie2,Vid, Jon1,Jon2,Har}.

Recently, it has attracted much attention to quantify the amount
of quantum coherence. In \cite{Bau}, the researchers established a
quantitative theory of coherence as a resource following the
approach that has been established for entanglement in
\cite{Ved2}. They introduced a rigorous framework for
quantification of coherence and proposed several measures of
coherence, which are based on the well-behaved metrics including
the $l_p$-norm, relative entropy, trace norm and fidelity.
Additional progress in this direction has been reported recently
in \cite{SXFL,XLF,MRS,MS,MC,RM,BDG,BD}.

However, as far as a finite number of coherence measures are
considered, the quantification of coherence is still in early
stages. This work is intended to contribute to a better
understanding of coherence. It presents a tool for build
infinitely many coherence measures. Our recipes shows how to build
all possible coherence measures for pure states (see sec. III).

By the tool of building coherence measures, we give the answer of
the question: suppose there is a pure coherent state
$|\psi\rangle=\sum_{i=1}^d\psi_i|i\rangle$ and we would like to
convert it into another pure coherent state
$|\phi\rangle=\sum_{i=1}^d\phi_i|i\rangle$ by incoherent
operations. Which is the greatest probability of success in such a
conversion? In \cite{Bau}, the authors provide a specific set of
Kraus operators that allow us--with finite probability--to
transform a pure state into another.  There, they remarked  that
this protocol may not be optimal.   In sec. IV, we provide  a
computation formula for the greatest probability and construct
explicitly an incoherent operation achieving the greatest
probability,  i.e., the optimal protocol.

\section{Preliminary}
Let ${\mathcal H}$ be a finite dimensional Hilbert space with
$d=\dim({\mathcal H})$. Fixing a particular basis
$\{|i\rangle\}_{i=1}^d$, we call all density operators (quantum
states) that are diagonal in this basis incoherent, and this set
of quantum states will be labelled by ${\mathcal I}$, all density
operators $\rho\in {\mathcal I}$ are of the form
$$\rho=\sum_{i=1}^d\lambda_i|i\rangle\langle i|.$$

{\it Incoherent operation---} A quantum operation $\Phi$ is a
trace-preserving completely positive linear map. By the classical
Kraus representation theorem, the quantum operation $\Phi$ can be
represented in an elegant form known as the operator-sum
representation. That is, $\Phi$ is an operation if and only if
there exist finite bounded linear operators $K_n$ satisfying
$\sum_n K_n^\dag K_n=I$ and $\Phi(\rho)=\sum_n K_n\rho
K_n^{\dag}$,   $I$ is the identity operator on ${\mathcal H}$.
From \cite{Bau}, the quantum operation $\Phi$ is incoherent if it
fulfils $K_n\rho K_n^\dag/Tr(K_n\rho K_n^\dag)\in {\mathcal I}$
for all $\rho\in {\mathcal I}$ and for all $n$. This definition
guarantees that in an overall quantum operation $\rho\mapsto
\sum_nK_n\rho K_n^\dag$, even if one does not have access to
individual outcomes $n$, no observer would conclude that coherence
has been generated from an incoherent state. It is easy to see
that a quantum operation is incoherent if and only if every column
of $K_n$ in the fixed basis $\{|i\rangle\}_{i=1}^d$ has at most
one nonzero entry.

Based on Baumgratz et al.'s suggestion \cite{Bau}, any proper
measure of coherence ${\mathcal C}$ must satisfy the following
axiomatic postulates.

(C1) The coherence measure vanishes on the set of incoherent
states, ${\mathcal C}(\rho)=0$ for all
$\rho\in{\mathcal I}$;

(C2a) Monotonicity under incoherent operation $\Phi$, ${\mathcal
C}(\Phi(\rho))\leq {\mathcal C}(\rho)$,

 or (C2b) Monotonicity
under selective measurements on average: $\sum_n p_n{\mathcal
C}(\rho_n)\leq {\mathcal C}(\rho)$, where $p_n=tr (K_n\rho
K_n^{\dag})$, $\rho_n=\frac {1}{p_n}K_n\rho K_n^{\dag}$, for all
$\{K_n\}$ with $\sum_n K_n^{\dag}K_n=I$ and $K_n\rho
K_n^\dag/Tr(K_n\rho K_n^\dag)\in {\mathcal I}$ for all $\rho\in
{\mathcal I}$;

(C3) Non-increasing under mixing of quantum states (convexity),
$${\mathcal C}(\sum_np_n\rho_n)\leq \sum_np_n{\mathcal
C}(\rho_n)$$ for any ensemble $\{p_n,\rho_n\}$.

 We remark that conditions (C2b) and (C3) imply condition (C2a).
And it has been recently shown that the coherence measure induced
by the fidelity satisfies (C2a), violates (C2b) \cite{SXFL}. For
the coherence measure induced by the trace norm, it is still not
known whether it satisfies criterion (C2b).

\section{building coherence measures}

The following  focuses on coherent measures for pure states and
extends these coherent measures  over the whole set of quantum
states. Our idea is originated  from \cite{vid2} which is devoted
to entanglement monotone. Similarly, we define  coherence monotone
to be any magnitude satisfying conditions (C2b) and (C3). From the
following Theorem 1 and Theorem 2, readers familiar with
entanglement theory will see, in the case of pure states, the $f$
considered in \cite{vid2} can derive a coherence monotone. While
the converse is not true. The key lies in that the entanglement
monotone is local unitary invariant, but the coherence monotone is
only invariant under some special unitary transformation (the
permutation of a diagonal unitary).

Let  $\Omega=\{{\bf x}=( x_1, x_2,\cdots,x_d)^t\mid \sum_{i=1}^d
x_i=1 \text{ and } x_i\geq 0 \}$, here $( x_1, x_2,\cdots,x_d)^t$
denotes the transpose of row vector $( x_1, x_2,\cdots,x_d)$. And
let $\pi$ be an arbitrary permutation of $\{1,2,\cdots, d\}$,
$P_{\pi}$ be the permutation matrix corresponding to $\pi$ which
is obtained by permuting the rows  of a $d\times d$ identity
matrix according to $\pi$. Given any nonnegative function
$f:\Omega\mapsto {\mathcal R}^{+}$ such that it is

\begin{itemize}
\item[$\bullet$] \begin{equation} \label{1} f(P_{\pi}(1,0 ,\cdots,
0)^t)=0,\end{equation} for every permutation $\pi$,

\item[$\bullet$] invariant under any permutation transformation
$P_{\pi}$, i.e.

\begin{equation} \label{2} f(P_{\pi}{\bf x})=f({\bf x}) \text{ for every } {\bf x}\in\Omega,  \end{equation}

\item[$\bullet$] concave, i.e. \begin{equation} \label{3}
f(\lambda {\bf x}+(1-\lambda){\bf y})\geq \lambda f({\bf
x})+(1-\lambda)f({\bf y})
\end{equation} for any $\lambda\in [0,1]$ and
${\bf x},{\bf y}\in\Omega$.
\end{itemize}

A coherence measure can be derived by defining it for pure states
(normalized vectors $|\psi\rangle
=(\psi_1, \psi_2, \cdots,
\psi_d)^t$ in the  fixed
basis $\{|i\rangle\}_{i=1}^d$ ) as
\begin{equation}\label{4} C_f(|\psi\rangle \langle
\psi|)=f((|\psi_1|^2, |\psi_2|^2, \cdots ,|\psi_d|^2)^t),
\end{equation} and by extending it over the whole set of density
matrices as
\begin{equation}\label{5} C_f(\rho)=\min _{p_j,\rho_j}\sum_j
p_jC_f(\rho_j),\end{equation} where the minimization is to be
performed over all the pure-state ensembles of $\rho$, i.e.,
$\rho=\sum_j p_j\rho_j$.

{\bf Theorem 1.} Any function $C_f$ satisfying (\ref{1})-(\ref{5})
is a coherence measure, i.e.,
\begin{equation} \text { Eqs. } (\ref{1})-(\ref{5}) \Rightarrow
C1,C2b,C3.\end{equation}

{\bf Proof:} For any $\rho\in {\mathcal I}$ and $\rho=\sum_i p_i
|i\rangle \langle i|$. From the definition of $C_f$ and
Eq.(\ref{1}), it follows that $C_f(\rho)\leq \sum_i
p_iC_f(|i\rangle \langle i|)=0$.

To verify C2b, we assume firstly that $\rho$ is a pure state
$|\psi\rangle \langle \psi|$. Here, it is  needed to use the
technique to deal with the acting of incoherent operations on pure
states developed in the proof of only if part of Theorem 1
\cite{BDG}. Considering that there is some subtle difference and
the completeness of the proof, we write it in detail. This
technique allows us to consider only the three dimensional case,
other cases can be treated similarly. Suppose
$|\psi\rangle=(\psi_1, \psi_2, \psi_3)^t$ and there is an
incoherent operation $\Phi$ with  Kraus operators $K_n$. Since,
for arbitrary permutation matrix $P_{\pi}$, $\{K_nP_{\pi}\}$ can
define an incoherent operation, we assume
\begin{equation}\label{9}|\psi_1|\geq |\psi_2|\geq |\psi_3|\end{equation} without loss of
generality. Denote $p_n=\|K_n |\psi\rangle\|^2$ and $\rho_n=\frac
1 {p_n}K_n |\psi\rangle\langle \psi|K_n^{\dag}$. Let
$k_j^{(n)}$($j=1,2,3$) be the nonzero element of $K_n$ at $j-th$
column (if there is no nonzero element in $j-th$ column, then
$k_j^{(n)}=0$). Suppose $k_j^{(n)}$ locates $f_n(j)-th$ row. Here,
$f_n(j)$ is a function that maps  $\{2,3\}$ to  $\{1,2,3\}$ with
the property that $1\leq f_n(j)\leq j$. Let
            $\delta_{s,t}=\left\{\begin{array}{cc}
            1,&s=t\\
            0,&s\neq  t\end{array}\right.$. Then there
is a permutation $\pi_n$ such that
\begin{equation}K_n=P_{\pi_n}\left(\begin{array}{ccc}
            k_1^{(n)} & \delta_{1,f_n(2)}k_2^{(n)} & \delta_{1,f_n(3)}k_3^{(n)}\\
            0         & \delta_{2,f_n(2)}k_2^{(n)} & \delta_{2,f_n(3)}k_3^{(n)}\\
            0         &    0                     &
            \delta_{3,f_n(3)}k_3^{(n)}\end{array}\right).\end{equation}
From $\sum_n K_n^{\dag}K_n=I$, we get that
\begin{equation}\label{6}\left\{\begin{array}{l}
\sum_n |k_j^{(n)}|^2=1, (j=1,2,3),\\
\sum_n \overline{k_1^{(n)}}\delta_{1,f_n(2)}k_2^{(n)}=0,\\
\sum_n \overline{k_1^{(n)}}\delta_{1,f_n(3)}k_3^{(n)}=0,\\
\sum_n
(\delta_{1,f_n(2)}\delta_{1,f_n(3)}+\delta_{2,f_n(2)}\delta_{2,f_n(3)})\overline{k_2^{(n)}}k_3^{(n)}=0.
\end{array}\right.\end{equation}
For $|\psi\rangle =(\psi_1,\psi_2,\psi_3)^t$, by a direct computation, one can get
\begin{equation}
 K_n|\psi\rangle=P_{\pi_n}\left(\begin{array}{c}
\phi^{(n)}_1\\
\phi^{(n)}_2\\
\phi^{(n)}_3\end{array}\right),\end{equation} here
\begin{equation}\left\{\begin{array}{l}
\phi^{(n)}_1=k_1^{(n)}\psi_1 + \delta_{1,f_n(2)}k_2^{(n)}\psi_2 +
\delta_{1,f_n(3)}k_3^{(n)}\psi_3,\\
\phi^{(n)}_2=\delta_{2,f_n(2)}k_2^{(n)}\psi_2+
\delta_{2,f_n(3)}k_3^{(n)}\psi_3,\\
\phi^{(n)}_3=\delta_{3,f_n(3)}k_3^{(n)}\psi_3.\end{array}\right.\end{equation}
Applying $\sum_n|\cdot|^2$ to above equations, we have
\begin{equation}\label{7}\left\{\begin{array}{l}
|\psi_1|^2 + \sum_n\delta_{1,f_n(2)}|k_2^{(n)}|^2|\psi_2|^2 \\
\ \ \ \ +
\sum_n\delta_{1,f_n(3)}|k_3^{(n)}|^2|\psi_3|^2=\sum_n|\phi^{(n)}_1 |^2,\\
\sum_n\delta_{2,f_n(2)}|k_2^{(n)}|^2|\psi_2|^2\\
\ \ \  \ +
\sum_n\delta_{2,f_n(3)}|k_3^{(n)}|^2|\psi_3|^2=\sum_n|\phi^{(n)}_2|^2 ,\\
\sum_n\delta_{3,f_n(3)}|k_3^{(n)}|^2|\psi_3|^2=\sum_n|\phi^{(n)}_3|^2.\end{array}\right.\end{equation}
Together with Eqs.(\ref{9}) and (\ref{6}), (\ref{7}) implies that
\begin{equation}\begin{array}{ll}
&((|\psi_1|^2,|\psi_2|^2,|\psi_3|^2)^t\\
\prec &(\sum_n|\phi^{(n)}_1 |^2, \sum_n|\phi^{(n)}_2|^2,
\sum_n|\phi^{(n)}_3|^2)^t.\end{array}\end{equation} Here
``$\prec$" is the majorization relation between vectors, the
definition and properties of which can be found in \cite{Bha}.
From Eqs.(\ref{2}) and (\ref{3}), it follows that
\begin{equation}\label{8}
\begin{array}{ll}
& \sum_n p_n C_f(\rho_n)\\ =& \sum_n p_n
f(P_{\pi_{n}}(\frac{|\phi^{(n)}_1 |^2} {p_n},\frac{|\phi^{(n)}_2
|^2}{p_n},\frac {|\phi^{(n)}_3
|^2}{p_n})^t)\\
= & \sum_n p_n f((\frac{|\phi^{(n)}_1 |^2}
{p_n},\frac{|\phi^{(n)}_2 |^2}{p_n},\frac {|\phi^{(n)}_3
|^2}{p_n})^t)\\
\leq & f((\sum_n |\phi^{(n)}_1 |^2,\sum_n |\phi^{(n)}_2 |^2, \sum_n
|\phi^{(n)}_3 |^2)^t) \\
\leq &
f((|\psi_1|^2,|\psi_2|^2,|\psi_3|^2)^t)=C_f(\rho).\end{array}\end{equation}
The last inequality is from \cite[Theorem II.3.3]{Bha}, that is,
any symmetric concave function is Schur-concave, i.e., $f({\bf
x})\geq f({\bf y})$ if ${\bf x}\prec {\bf y}$.

Suppose now that $\rho$ is any mixed state. Let $\rho=\sum_i
q_{i}\sigma_{i}$ is an optimal pure-state ensemble with
$C_f(\rho)=\sum_i q_{i}C_f(\sigma_{i})$. Then
\begin{equation}\begin{array}{ll}
& \sum_n p_n C_f(\rho_n)=\sum_n p_n C_f(\frac{\sum_i q_i
K_n\sigma_i K_n^{\dag}}{p_n})\\
\leq & \sum_n p_n \sum_i q_i\frac{tr(K_n\sigma_i K_n^{\dag})}{p_n}
C_f(  \frac{K_n\sigma_i K_n^{\dag}}{tr(K_n\sigma_i K_n^{\dag})})\\
= & \sum_i q_i(\sum_n tr(K_n\sigma_i K_n^{\dag})C_f(  \frac{K_n\sigma_i K_n^{\dag}}{tr(K_n\sigma_i K_n^{\dag})}))\\
\leq &\sum_i q_iC_f(\sigma_i)=C_f(\rho).
\end{array}\end{equation} Two inequalities follow from Eq.(\ref{5}) and Eq.(\ref{8}), respectively.

Now we  prove C3 holds true. Let $\rho=\sum_i p_i\rho_i$ be any
ensemble of $\rho$. And let $\rho_i=\sum_j q_{ij}\rho_{ij}$ be an
optimal pure-state ensemble of $\rho_i$, i.e., $C_f(\rho_i)=\sum_j
q_{ij}C_f(\rho_{ij})$. Then
\begin{equation} \begin{array}{ll}
& C_f(\rho)=C_f(\sum_ip_i\sum_jq_{ij}\rho_{ij})\\
=& C_f(\sum_{ij}p_iq_{ij}\rho_{ij})\leq
\sum_{ij}p_iq_{ij}C_f(\rho_{ij})\\
= & \sum_i p_i C_f(\rho_i).\end{array}\end{equation} The
inequality follows from Eq.(5).\hspace{0.1in}$\square$

As examples of coherence measures built using Theorem 1 consider,
for any nonnegative function $\widehat{f}(x)$ concave in the
interval $x\in [0, 1]$ with $\widehat{f}(0)=\widehat{f}(1)=0$, the
function $f : \mathcal R^d\rightarrow {\mathcal R}^+$  defined by
$f((x_1,x_2,\cdots,x_d)^t)=\sum_i \widehat{f}(x_i)$. Then $f$
satisfies Eqs.(\ref{1})-(\ref{3}). Taking
\begin{equation}\widehat{f}(x) = -x \log_2 x,\end{equation} we can
induce the coherence measure ${\mathcal C}_f$ which is identical
with the relative entropy coherence measure \cite{Bau} on pure
states. Generally speaking, they are different on mixed states.
Choosing
\begin{equation}f((x_1,x_2,\cdots,x_d)^t)=(\sum_{i=1}^d \sqrt{x_i})^2-1,\end{equation} we
can easily check  that the coherence measure ${\mathcal C}_f$ is
identical with $l_1$-norm coherence measure \cite{Bau} on pure
states. They are indeed different on mixed states. It is well
known that $\alpha$-entropy are used to measure the uncertainty.
We can also define $\alpha$-entropy coherence measure. Let
\begin{equation}f((x_1,x_2,\cdots,x_d)^t)=\frac
1{1-\alpha}\log_2\sum_{i=1}^dx_i^{\alpha},\hspace{0.1in} 0<\alpha
<1.\end{equation} It follows from the fact that the logarithm is a
concave, non-decreasing function and therefore preserves
concavity, that the $\alpha$-entropy is a concave function. The
Eqs.(\ref{1})(\ref{2}) are easy to check. Consequently,  Theorem 1
can be applied directly to prove that the $\alpha$-entropy can
derive a coherence measure.

In the following, we will show that one can construct
any coherence measure for pure states by our strategy.

{\bf Theorem 2.} The restriction of any coherence measure
(satisfying C1,C2b and C3) to pure states  can be derived by a
function $f:\Omega\rightarrow \mathcal R^+$ satisfying
Eqs.(\ref{1})-(\ref{3}).

{\bf Proof:} Let $\mu$ be an arbitrary coherence measure and let $U$ be a
 diagonal unitary matrix. From the
monotonicity of coherence measures under incoherent operations, it
is evident that $\mu(U\rho U^\dag)\leq
\mu(\rho)$. Symmetrically, one can see that $\mu(\rho)=
\mu(U^\dag(U\rho U^\dag)U) \leq
\mu(U\rho U^\dag)$. So $\mu(U\rho U^\dag)=
\mu(\rho)$.
Define $f:\Omega\rightarrow \mathcal R^+$ by
$f((x_1,x_2,\cdots,x_d)^t)=\mu(|\psi\rangle\langle \psi|)$, where
$|\psi\rangle=\sum_{i=1}^d \sqrt{x_i}|i\rangle$.  For any pure
state $|\psi\rangle=\sum_i \psi_i|i\rangle$, there exists diagonal
unitary matrix $U$ such that $U|\psi\rangle=\sum_i
|\psi_i||i\rangle$. It follows that
\begin{equation}\mu(|\psi\rangle\langle
\psi|)=\mu(U|\psi\rangle\langle\psi|
U^\dag)=f((|\psi_1|^2,|\psi_2|^2,\cdots,|\psi_d|^2)^t).\end{equation}

In the following, we will check that $f$ satisfies
Eqs.(\ref{1})-(\ref{3}). The Eq.(\ref{1}) follows from C1. Let
$\pi$ be a permutation of $\{1,2,\cdots, d\}$, by the definition
of $f$, we have
\begin{equation}f((x_{\pi(1)},x_{\pi(2)},\cdots,x_{\pi
(d)})^t)=\mu(P_{\pi}|\psi\rangle\langle \psi|P_{\pi}).\end{equation}
By the same argument as the diagonal unitary matrix case, one can obtain
\begin{equation}\mu(P_{\pi}|\psi\rangle\langle \psi|P_{\pi})=
\mu(|\psi\rangle\langle \psi|).\end{equation} This implies that
$f$ is invariant under any permutation transformation. To prove
Eq.(\ref{3}), for $${\bf x}=(x_1,x_2, \cdots, x_d)^t,$$ $${\bf
y}=(y_1,y_2, \cdots, y_d)^t\in\Omega$$ and $\lambda\in[0,1]$, we
define $$\begin{array}{l} K_1=\text{diag}(\frac{\sqrt{\lambda
x_1}}{\sqrt{\lambda x_1+(1-\lambda) y_1}}, \frac{\sqrt{\lambda
x_2}}{\sqrt{\lambda x_2+(1-\lambda)y_2}},\\
\ \ \ \ \ \ \ \ \ \ \ \ \ \ \ \ \ \cdots, \frac{\sqrt{\lambda
x_d}}{\sqrt{\lambda x_d+(1-\lambda)y_d}}),\end{array}$$
$$\begin{array}{l} K_2=\text{ diag}(\frac{\sqrt{(1-\lambda)y_1}}{\sqrt{\lambda
x_1+(1-\lambda)y_1}}, \frac{\sqrt{(1-\lambda)y_2}}{\sqrt{\lambda
x_2+(1-\lambda)y_2}},\\ \ \ \ \ \ \ \ \ \ \ \ \ \ \ \ \ \ \cdots,
\frac{\sqrt{(1-\lambda)y_d}}{\sqrt{\lambda x_d+(1-\lambda
)y_d}}).\end{array}$$ It is easy to check that
$\Phi(\cdot)=K_1\cdot K_1^{\dag}+K_2\cdot K_2^{\dag}$ is an
incoherent operation. And
\begin{equation}K_1\sum_i\sqrt{\lambda x_i+(1-\lambda)y_i}|i\rangle=\sqrt{\lambda }\sum_i
\sqrt{x_i}|i\rangle,\end{equation}
\begin{equation}K_2\sum_i\sqrt{\lambda x_i+(1-\lambda )y_i}|i\rangle=\sqrt{1-\lambda}\sum_i
\sqrt{y_i}|i\rangle.\end{equation} From (C2b), we get that
\begin{equation}\begin{array}{ll}
&\lambda \mu( \sum_i \sqrt{x_i}|i\rangle)+(1-\lambda) \mu( \sum_i
\sqrt{y_i}|i\rangle) \\
\leq &  \mu
(\sum_i\sqrt{\lambda x_i+(1-\lambda)y_i}|i\rangle).\end{array}\end{equation} That is $f(\lambda
{\bf x}+(1-\lambda ){\bf y})\geq \lambda f({\bf x})+(1-\lambda)f({\bf
y})$, i.e., $f$ is concave.
 $\square$

\section{optimal conversion for coherent states}
The section is devoted to the optimal conversion probability in a
single-copy scenario. In \cite{Vid}, an optimal local conversion
strategy between any two pure entangled  states of a bipartite
system is presented . In \cite{Fer}, Brand\~{a}o and  Gour have
proposed a general framework to analyse the conversion in the
asymptotic limit and shown that a quantum resource theory is
asymptotically reversible if its set of
allowed operations is maximal. 

For pure states $|\psi\rangle=\sum_{i=1}^d\psi_i|i\rangle,
|\phi\rangle=\sum_{i=1}^d\phi_i|i\rangle$, we can assume that
$|\psi_1|\geq|\psi_2|\geq \cdots\geq|\psi_d|$ and
$|\phi_1|\geq|\phi_2|\geq \cdots\geq|\phi_d|$. Indeed, in general
case, there exist two permutations $\pi,\sigma$ of
$\{1,2,\cdots,d\}$ such that $|\psi_{\pi(1)}|\geq
|\psi_{\pi(2)}|\geq \cdots\geq |\psi_{\pi(d)}|$ and
$|\phi_{\sigma(1)}|\geq |\phi_{\sigma(2)}|\geq \cdots\geq
|\phi_{\sigma(d)}|$.  Let $U=P_{\pi}$ and $V=P_{\sigma}$, here
$P_{\pi}$ and $P_{\sigma}$ are permutation matrices corresponding
to $\pi$ and $\sigma$, respectively. Note that $U|\psi\rangle
\xrightarrow{ICO}V|\phi\rangle\Leftrightarrow|\psi\rangle
\xrightarrow{ICO}|\phi\rangle$, here $|\psi\rangle
\xrightarrow{ICO}|\phi\rangle$ indicates that
$|\psi\rangle\langle\psi|$ is transformed to
$|\phi\rangle\langle\phi|$ by an incoherent operation. Therefore
we can replace $|\psi\rangle$ and $|\phi\rangle$ by
$U|\psi\rangle$ and $V|\phi\rangle$. Furthermore,
$P(|\psi\rangle\xrightarrow{ICO}|\phi\rangle)=P(U|\psi\rangle\xrightarrow{ICO}|V\phi\rangle)$.
Here $P(|\psi\rangle\xrightarrow{ICO}|\phi\rangle)$ denotes the
greatest probability  of success under incoherent operations
transferring $|\psi\rangle$ to $|\phi\rangle$.


{\bf Theorem 3.}
$P(|\psi\rangle\xrightarrow{ICO}|\phi\rangle)=\min_{l\in[1,d]}
\frac {\sum_{i=l}^d |\psi_i|^2}{\sum_{i=l}^d |\phi_i|^2}$.

{\bf Proof:} We will show the  equation by verifying that
$P(|\psi\rangle\xrightarrow{ICO}|\phi\rangle)\leq \frac
{\sum_{i=l}^d |\psi_i|^2}{\sum_{i=l}^d |\phi_i|^2}$ for each $l$
and giving an optimal incoherent operation.

In the case of $l=1$, it is trivial, since
$P(|\psi\rangle\xrightarrow{ICO}|\phi\rangle)\leq 1=\frac
{\sum_{i=1}^d |\psi_i|^2}{\sum_{i=1}^d |\phi_i|^2}$. For the case
of  $l\neq 1$, define $f_l:\Omega\rightarrow {\mathcal R}^+$ by
$f_l((x_1,x_2,\cdots,x_d)^t)=\sum_{i=l}^d x_i^{\downarrow}$, here
$(x_1^{\downarrow},x_2^{\downarrow},\cdots,x_d^{\downarrow})^t$ is
the vector obtained by rearranging the coordinates of
$(x_1,x_2,\cdots,x_d)^t$ in the decreasing order. We firstly check
that $f_l$ satisfies Eqs.(\ref{1}-\ref{3}). Since $l\geq 2$,
$f_l((1,0,\cdots, 0)^t)=\sum_{i=l}^d 0=0$. By the definition of
$f_l$, it is clear that $f_l$ is invariant under any permutation
transformation. $f_l$ is a concave function follows from the Ky
Fan's maximum principle \cite[Page 24]{Bha}.

From Theorem 1, it follows that it can derive a coherence measure
$C_{f_l}$.  From the C2b and neglecting positive contributions
coming from unsuccessful conversions, it follows that
\begin{equation}P(|\psi\rangle\xrightarrow{ICO}|\phi\rangle)C_{f_l}(|\phi\rangle\langle\phi|)\leq
C_{f_l}(|\psi\rangle\langle\psi|).\end{equation}

Now we give the optimal incoherent operation. The strategy is
borrowed from \cite{Vid} which consider similar problem in the
frame of entanglement. The key difference lies in replacing the
Nielsen Theorem by the corresponding part about coherent
transformation which is recently proposed in \cite{BDG}. For the
convenience of readers, we also provide the details.

We divide into two steps. In the first, by using the result in
\cite{BDG}, we will show that an incoherent operation transfer the
initial state $|\psi\rangle$ into a temporary pure state
$|\gamma\rangle$ with certainty. Secondly, $|\gamma\rangle$ is
transfered into $|\phi\rangle$ by mean of incoherent operation
with the probability $\min_{l\in[1,d]} \frac {\sum_{i=l}^d
|\psi_i|^2}{\sum_{i=l}^d |\phi_i|^2}$.

Let $l_1$ be the smallest integer in $[1,d]$ such that
\begin{equation} \frac {\sum_{i=l_1}^d
|\psi_i|^2}{\sum_{i=l_1}^d |\phi_i|^2}=\min_{l\in[1,d]} \frac
{\sum_{i=l}^d |\psi_i|^2}{\sum_{i=l}^d |\phi_i|^2}\equiv
r_1.\end{equation} It may happen that $l_1=r_1=1$. If not, it
follows from the equivalence
\begin{equation}\frac{a}{b}<\frac{a+c}{b+d}\Leftrightarrow\frac{a}{b}<\frac{c}{d}(a,b,c,d>0)\end{equation}
that for any integer $k\in[1,l_{1}-1]$ such that $ \frac
{\sum_{i=k}^{l_1-1} |\psi_i|^2}{\sum_{i=k}^{l_1-1}
|\phi_i|^2}>r_1$. Let us then define $l_2$ as the smallest integer
$\in [1,l_1-1]$ such that
\begin{equation} r_2=\frac {\sum_{i=l_2}^{l_1-1}
|\psi_i|^2}{\sum_{i=l_2}^{l_1-1} |\phi_i|^2}=\min_{l\in[1,l_1-1]}
\frac {\sum_{i=l}^{l_1-1} |\psi_i|^2}{\sum_{i=l}^{l_1-1}
|\phi_i|^2}\quad (>r_1).\end{equation} Repeating this process
until $l_k=1$ for some $k$, we obtain s series of $k+1$ integers
$l_0>l_1>l_2>\cdots l_k$ ($l_0=d+1$), and $k$ positive numbers $0<
r_1 <r_2<\cdots <r_k$, by the means of which we define our
temporary (normalized) state \begin{equation}\begin{array}{l}
|\gamma\rangle=\sum_{i=1}^d \gamma _i|i\rangle, \text{ where }\\
\gamma_i=\sqrt{r_j}\phi_i \quad \text{ if } i\in[l_j,l_{j-1}-1],
1\leq j\leq k\end{array}\end{equation} i.e.,
\begin{equation}\overrightarrow{\gamma}=\left(\begin{array}{c}
\sqrt{r_k}\left(\begin{array}{c}\phi_{l_k}\\ \vdots\\
\phi_{l_{k-1}-1}\end{array}\right)\\
\vdots\\
\sqrt{r_2}\left(\begin{array}{c}\phi_{l_2}\\ \vdots\\
\phi_{l_{1}-1}\end{array}\right)\\
\sqrt{r_1}\left(\begin{array}{c}\phi_{l_1}\\ \vdots\\
\phi_{l_{0}-1}\end{array}\right)\\
\end{array}\right)\end{equation}
From the construction, it follows that
\begin{equation}\sum_{i=k}^d |\psi_i|^2\geq \sum_{i=k}^d
|\gamma_i|^2 \quad \forall k\in[1,d],\end{equation} which is
equivalent that $\sum_{i=1}^k |\psi_i|^2\leq \sum_{i=1}^k
|\gamma_i| ^2\quad \forall k\in[1,d]$. By \cite[Theorem 1]{BDG},
there exists an  incoherent operation transferring $|\psi\rangle$
into $|\gamma\rangle$ with certainty.

Define the positive operator $M: {\mathbb C}^d \rightarrow
{\mathbb C}^d $ by
\begin{equation}M=\left(\begin{array}{cccc}M_k & & &\\& \ddots & &\\ & & M_2 &\\
&&& M_1\end{array}\right),\end{equation}where
\begin{equation}M_j=\sqrt{\frac{r_1}{r_j}}I_{[l_{j-1}-l_j]}, \quad
j=1,2,\cdots,k,\end{equation} is proportional to the identity in
$(l_{j-1}-l_j)$-dimensional subspace of $\mathbb C ^d$. So that
$M,\sqrt{I-M^2}$ define an incoherent operation satisfying
$M|\gamma\rangle=\sqrt{r_1}|\phi\rangle$. $\square$

At the end of the section, we  consider two alternative scenarios where Theorem 3 can be
applied. At first, we consider the greatest probability  of copies of state
$|\phi\rangle$ transferred from $|\psi\rangle$, denote it by
$m_{|\psi\rangle\rightarrow |\phi\rangle}^{max}$, i.e., $m_{|\psi\rangle\rightarrow |\phi\rangle}^{max}
=\max_nP(|\psi\rangle \xrightarrow{ICO}|\phi\rangle^{\otimes n})$.   In general, this cannot be obtained
by Theorem 3  directly. However, there are circumstances in which
$m_{|\psi\rangle\rightarrow
|\phi\rangle}^{max}=P(|\psi\rangle\xrightarrow{ICO}|\phi\rangle)$.
Indeed, let $n_{|\psi\rangle}$  denote the number of nonvanishing
coefficients of the entangled state $|\psi\rangle$, and recall
that $n_{|\psi\rangle^{\otimes N}}=n_{|\psi\rangle}^N$. Then,
\begin{equation}n_{|\psi\rangle} < n_{|\phi\rangle}^2\Rightarrow P(|\psi\rangle\xrightarrow{ICO}|\phi\rangle^{\otimes
N})=0 \quad N\geq 2\end{equation} implies that
$m_{|\psi\rangle\rightarrow
|\phi\rangle}^{max}=P(|\psi\rangle\xrightarrow{ICO}|\phi\rangle)$
when $n_{|\psi\rangle} < n_{|\phi\rangle}^2$.
Secondly, from Theorem 3, we also get that one can often extract more
coherence from two copies of a given state $|\psi\rangle$, i.e.,
$|\psi\rangle^{\otimes 2}$, than twice what they can obtain from
one single copy $|\psi\rangle$. For example, $|\psi\rangle=(\frac
1 {\sqrt 2})(|1\rangle+|2\rangle)$ and $|\phi\rangle=(\frac 1
{\sqrt 3})(|1\rangle+|2\rangle+|3\rangle)$. Then
$1=P(|\psi\rangle^{\otimes
2}\xrightarrow{ICO}|\phi\rangle)>P(|\psi\rangle\xrightarrow{ICO}|\phi\rangle)=0$.


\section{Conclusion}
This paper is focused on quantification of coherence. We have
provided a tool to build an important class of coherence measures
which cover the relative entropy measure for pure states, the
$l_1$-norm measure for pure states, and the $\alpha$-entropy
measure. Furthermore, any
 coherence measure on pure coherent states can be
constructed in this way. Using a set of coherence measure and
constructing the optimal conversion, we give the explicit
expression of the greatest probability
$P(|\psi\rangle\xrightarrow{ICO}|\phi\rangle)$ of success in the
conversion of given  states  $|\psi\rangle$ and $|\phi\rangle$
under incoherent operations.

\section{Acknowledgement}

The authors thank referees for thorough reading of our paper and valuable comments.
This work was completed while the authors were visiting the IQC of
the University of Waterloo and Department of Mathematics and
Statistics of the University of Guelph  during the academic year
2014-2015 under the support of China Scholarship Council. We thank
Professor David W. Kribs and Professor Bei Zeng for their
hospitality. This work is partially supported by the Natural
Science Foundation of China (No. 11001230),the Natural Science
Foundation of Fujian (2013J01022, 2014J01024) and and Program for
the Outstanding Innovative Teams of Higher Learning Institutions
of Shanxi.


\begin{thebibliography} {99}

\bibitem{Nielsen} M. A. Nielsen and I. L. Chuang (2000), {\it Quantum Computation
and Quantum information}, Cambridge University Press (Cambridge).


\bibitem{Ben1} C. H. Bennett et al. (1996), {\it Mixed state entanglement and quantum error correction},
Phys. Rev. A, 54, 3824-3851.

\bibitem{Ben2} C. H. Bennett et al. (1996), {\it Concentrating partial entanglement by local operations},
Phys. Rev. A, 53, 2046-2052.

\bibitem{Ben3} C. H. Bennett et al. (1996), {\it Purification of noisy
entanglement and faithful teleportation via noisy channels}, Phys.
Rev. Lett., 76, 722-725.

\bibitem{Ved1} V. Vedral et al., (1997), {\it Quantifying entanglement}, Phys. Rev. Lett., 78,
2275-2279.

\bibitem{Ved2} V. Vedral and M. B. Plenio. (1998), {\it Entanglement measures and purification procedures},
Phys. Rev. A, 57, 1619-1633.


\bibitem{vid2} G. Vidal (2000),  {\it Entanglement monotones},
 J. Mod. Opt., 47,  355-376.

\bibitem{Nie2} M. A. Nielsen (1999), {\it Conditions for a class of entanglement transformations},
Phys. Rev. Lett., 83, 436-439.

\bibitem{Vid} G. Vidal (1999), {\it Entanglement of pure states for a single copy}, Phys. Rev. Lett., 83, 1046-1049.


\bibitem{Jon1} D. Jonathan and M. B. Plenio (1999), {\it Minimal conditions for local pure-state entanglement manipulation},
Phys. Rev. Lett., 83, 1455-1458.

\bibitem{Jon2} D. Jonathan and M. B. Plenio (1999), {\it Entanglement-Assisted local manipulation of pure quantum states},
Phys. Rev. Lett., 83, 3566-3569.

\bibitem{Har} L. Hardy (1999), {\it Method of areas for manipulating the
 entanglement properties of one copy of a two-particle pure entangled state},
Phys. Rev. A, 60, 1912-1917.

\bibitem{Bau} T. Baumgratz, M. Cramer, and M. B. Plenio (2014), {\it Quantifying coherence},
 Phys. Rev. Lett., 113, 140401.
.

\bibitem{SXFL} L. H. Shao, Z. J. Xi, H. Fan and Y. M.
Li (2015), {\it Fidelity and trace-norm distances for quantifying
coherence}, Phys. Rev. A, 91, 042120.

\bibitem{XLF} Z. Xi, Y. Li and H. Fan (2015), {\it Quantum coherence and correlations in quantum system},
Sci. Rep., 5. 10922.

\bibitem{MRS} I. Marvian, Robert W. Spekkens (2014), {\it Modes of asymmetry: the application of harmonic
analysis to symmetric quantum dynamics and quantum reference
frames},  Phys. Rev. A, 90, 062110.

\bibitem{MS} Iman Marvian and RobertW. Spekkens (2014), {\it Extending Noether¡¯s theorem by quantifying the asymmetry
of quantum states}, Nat. Commun., 5, 3821.

\bibitem{MC} A. Monras, A.Ch\c{e}ci\'{n}ska and A. Ekert (2014) , {\it Witnessing quantum coherence in the presence of noise},
  New J. Phys., 16, 063041.

\bibitem{RM} \'{A}. Rivas and M. M\"{u}ller (2015), {\it Quantifying spatial correlations of general quantum dynamics},
 New J. Phys., 17, 062001.

\bibitem{BDG} S. P.  Du, Z. F. Bai and Y. Guo (2015), {\it Conditions for coherence transformations
 under incoherent operations
},  Phys. Rev. A, 91, 052120.

\bibitem{BD} S. P.  Du, Z. F. Bai (2015),  {\it The Wigner--Yanase information can increase
under phase sensitive incoherent operations}, Annals of Phys.,
359, 136-140.


\bibitem{Bha} R. Bhatia (1997), {\it Matrix Analysis}, Springer-Verlag, (New York), .



\bibitem{Fer} Fernando G.S.L. Brand\~{a}o and Gilad Gour, arXiv:1502.03139v1.

\end{thebibliography}
\end{document}